# Distributed Clustering Algorithm for Spatial Data Mining

Malika Bendechache[#1], M-Tahar Kechadi[#2]

[#]School of Computer Science & Informatics, University College Dublin
Belfield, Dublin 04, Ireland
[1]malika.bendechache@ucdconnect.ie
[2]Tahar.kechadi@ucd.ie

*Abstract*— Distributed data mining techniques and mainly distributed clustering are widely used in the last decade because they deal with very large and heterogeneous datasets which cannot be gathered centrally. Current distributed clustering approaches are normally generating global models by aggregating local results that are obtained on each site. While this approach mines the datasets on their locations the aggregation phase is complex, which may produce incorrect and ambiguous global clusters and therefore incorrect knowledge. In this paper we propose a new clustering approach for very large spatial datasets that are heterogeneous and distributed. The approach is based on K-means Algorithm but it generates the number of global clusters dynamically. Moreover, this approach uses an elaborated aggregation phase. The aggregation phase is designed in such a way that the overall process is efficient in time and memory allocation. Preliminary results show that the proposed approach produces high quality results and scales up well. We also compared it to two popular clustering algorithms and show that this approach is much more efficient.

## I. INTRODUCTION

Across a wide variety of fields, datasets are being collected and accumulated at a dramatic pace and massive amounts of data that are being gathered are stored in different sites. In this context, data mining (DM) techniques have become necessary for extracting useful knowledge from the rapidly growing large and multi-dimensional datasets [1]. In order to cope with large volumes of data, researchers have developed parallel versions of the sequential DM algorithms [2]. These parallel versions may help to speedup intensive computations, but they introduce significant communication overhead, which make them inefficient. To reduce the communication overheads distributed data mining (DDM) approaches that consist of two main steps are proposed. As the data is usually distributed the first phase consists of executing the mining process on local datasets on each node to create local results. These local results will be aggregated to build global ones. Therefore the efficiency of any DDM algorithm depends closely on the efficiency of its aggregation phase. In this context, distributed data mining (DDM) techniques with efficient aggregation phase have become necessary for analysing these large and multi-dimensional datasets. Moreover, DDM is more appropriate for large-scale distributed platforms, such as clusters and Grids [3], where datasets are often geographically distributed and owned by different organisations. Many DDM methods such as distributed association rules and distributed classification [4], [5], [6], [7], [8], [9] have been proposed and developed in the last few years. However, only a few research concerns distributed clustering for analysing large, heterogeneous and distributed datasets. Recent researches [10], [11], [12], [13] have proposed distributed clustering approaches based on the same 2-step process: perform partial analysis on local data at individual sites and then send them to a central site to generate global models by aggregating the local results. In this paper, we propose a distributed clustering approach based on the same 2-step process, however, it reduces significantly the amount of information exchanged during the aggregation phase, generates automatically the correct number of clusters, and also it can use any clustering algorithm to perform the analysis on local datasets. A case study of an efficient aggregation phase has been developed on special datasets and proven to be very efficient; the data exchanged is reduced by more than 98% of the original datasets [15].

The rest of this paper is organised as follows: In the next section we will give an overview of distributed data mining and discuss the limitations of traditional techniques. Then we will present and discuss our approach in Section 3. Section 4 presents the implementation of the approach and we discuss experimental results in Section 5. Finally, we conclude in Section 6.

## II. DISTRIBUTED DATA MINING

Existing DDM techniques consist of two main phases: 1) performing partial analysis on local data at individual sites and 2) generating global models by aggregating the local results. These two steps are not independent since naive approaches to local analysis may produce incorrect and ambiguous global data models. In order to take advantage of mined knowledge at different locations, DDM should have a view of the knowledge that not only facilitates their integration but also minimises the effect of the local results on the global models. Briefly, an efficient management of distributed knowledge is one of the key factors affecting the outputs of these techniques.

Moreover, the data that will be collected in different locations using different instruments may have different formats, features, and quality. Traditional centralised data mining techniques do not consider all the issues of data-driven applications such as scalability in both response time and accuracy of solutions, distribution and heterogeneity [8], [16].

Some DDM approaches are based on ensemble learning, which uses various techniques to aggregate the results [11],

among the most cited in the literature: majority voting, weighted voting, and stacking [17], [18]. Some approaches are well suited to be performed on distributed platforms. For instance, the incremental algorithms for discovering spatio-temporal patterns by decomposing the search space into a hierarchical structure, addressing its application to multi-granular spatial data can be very easily optimised on hierarchical distributed system topology. From the literature, two categories of techniques are used: parallel techniques that often require dedicated machines and tools for communication between parallel processes which are very expensive, and techniques based on aggregation, which proceed with a purely distributed, either on the data based models or on the execution platforms [7], [12]. However, the amount of data continues to increase in recent years, as such, the majority of existing data mining techniques are not performing well as they suffers from the scalability issue. This becomes a very critical issue in recent years. Many solutions have been proposed so far. They are generally based on small improvements to fit a particular data at hand.

Clustering is one of the fundamental techniques in data mining. It groups data objects based on information found in the data that describes the objects and their relationships. The goal is to optimise similarity measure within a cluster and the dissimilarities between clusters in order to identify interesting structures/patterns/models in the data [12]. The two main categories of clustering are partitioning and hierarchical. Different elaborated taxonomies of existing clustering algorithms are given in the literature and many distributed clustering versions based on these algorithms have been proposed in [12], [20], [21], [22], [23], [24], [25], etc. Parallel clustering algorithms are classified into two sub-categories. The first consists of methods requiring multiple rounds of message passing. They require a significant amount of synchronization. The second sub-category consists of methods that build local clustering models and send them to a central site to build global models [15]. In [20] and [24], message-passing versions of the widely used k-means algorithm were proposed. In [21] and [25], the authors dealt with the parallelization of the DBSCAN density based clustering algorithm. In [22] a parallel message passing version of the BIRCH algorithm was presented. A parallel version of a hierarchical clustering algorithm, called MPC for Message Passing Clustering, which is especially dedicated to Microarray data, was introduced in [23]. Most of the parallel approaches need either multiple synchronization constraints between processes or a global view of the dataset, or both [12].

Both partitioning and hierarchical categories have some weaknesses. For the partitioning class, the k-means algorithm requires the number of clusters to be fixed in advance, while in the majority of cases K is not known, furthermore hierarchical clustering algorithms have overcome this limitation, but they must define the stopping conditions for clustering decomposition, which are not straightforward.

III. SPATIAL DISTRIBUTED CLUSTERING

The proposed distributed approach follows the traditional two-step strategy; 1) it first generates local clusters on each sub-dataset that is assigned to a given processing node, 2) these local clusters are aggregated to form global ones. This approach is developed for clustering spatial datasets. The local clustering algorithm can be any clustering algorithm. For sake of clarity it is chosen to be K-Means executed with a given ($K_i$) which can be different for each node (see Figure 1). $K_i$ should be chosen to be big enough to identify all clusters in the local datasets.

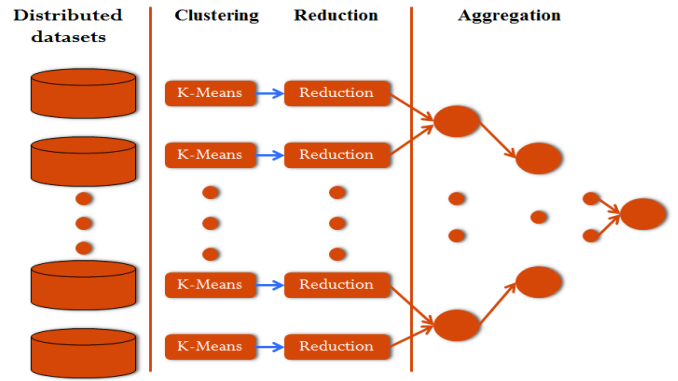

Fig 1. Overview of the Proposed Approach

After generating local results, each node compares its local clusters with its neighbours' clusters. Some of the nodes, called leader, will be elected to merge local clusters to form larger clusters using the overlay technique. These leaders are elected according to some conditions such as their capacity, their processing power, etc. The process of merging clusters will continue until we reach the root node. The root node will contain the global clusters (models).

During the second phase, communicating the local clusters to the leaders may generate huge overhead. Therefore, the objective is to minimise the data communication and computational time, while getting accurate global results. In fact our approach minimises the overheads due to the data exchange. Therefore instead of exchanging the whole data (whole clusters) between nodes (local nodes and leaders), we first proceed by reducing the data that represent a cluster. The size of this new data cluster is much smaller that the initial one. This process is carried out on each local node.

There are many data reduction techniques proposed in the literature. Many of them are focusing only in dataset size i.e., they try to reduce the storage of the data without paying attention to the knowledge behind this data. In [**26**], an efficient reduction technique has been proposed; it is based on density-based clustering algorithms. Each cluster consists of its representatives. However, selecting representatives is still a challenge in terms of quality and size. We can choose, for instance, medoids points, core points, or even specific core points [10] as representatives [15].

We focus on the shape and the density of the clusters. The shape of a cluster is represented by its boundary points (called contour) (see *Fig 2*). Many algorithms for extracting the boundaries from a cluster can be found in the literature [27], [28], [29], [30], [31]. We used the algorithm proposed in [32] which is based on Triangulation to generate the cluster boundaries. It is an efficient algorithm for constructing non-convex boundaries. The algorithm is able to accurately characterise the shape of a wide range of different point distributions and densities with a reasonable complexity of *O(n log n)*.

The boundaries of the clusters represents the new dataset, and they are much more smaller than the original datasets. So the boundaries of the clusters will become the local results at each node in the system. These local results are sent to the leaders following a tree topology. The global results will be located at the root of the tree.

## IV. IMPLEMENTED APPROACH

### A. Distributed Dynamic Clustering Algorithm ($D^2CA$)

In the first phase, called the *parallel phase*, the local clustering is performed using the K-means algorithm. Each node ($d_i$) executes K-means on its local dataset to produce $K_i$ local clusters. Once all the local clusters are determined, we calculate their contours. These contours will be used as representatives of their corresponding clusters. The second phase of the technique consists of exchanging the contours of each node with its neighbourhood nodes. This will allow us to see if there are any overlapping contours (clusters).

In the third step each leader attempts to merge overlapping contours of its group. The leaders are elected among nodes of each group. Therefore, each leader generates new contours (new clusters). We repeat the second and third steps till we reach root node. The sub-clusters aggregation is done following a tree structure and the global results are located in the top level of the tree (root node).

As in all clustering algorithms, the expected large variability in clusters shapes and densities is an issue. However, as we will show in the next section, the algorithm used for generating the cluster's contour is efficient to detect well-separated clusters with any shapes. Moreover $D^2CA$ determines also dynamically the number of the clusters without a priori knowledge about the data or an estimation process of the number of the clusters. In the following we will describe the main features and the requirements of the algorithm and its environment.

The nodes of the distributed computing system are organised following a tree topology.
1) Each node is allocated a dataset representing a portion of the scene or of the overall dataset.
2) Each leaf node ($n_i$) executes the K-means algorithm with $K_i$ parameter on its local data.
3) Neighbouring nodes must share their clusters to form even larger clusters using the overlay technique.
4) The results must reside in the father node (called ancestor).
5) Repeat 3 and 4 until reaching the root node.

In the following we give a pseudo-code of the algorithm:

**Algorithm 1**: *Distributed Dynamic Clustering Algorithm* ($D^2CA$)
**Input** : $X_i$: Dataset Fragment, $K_i$: Number of sub-clusters for $Node_i$, D: tree degree.
**Output**: $K_g$: Global Clusters (global results)
level = treeheight;
1) K-means($X_i$, $K_i$);
   // $Node_i$ executes K-Means algorithm locally.
2) Contour(K_i);
   // Node-i executes Contour algorithm to generate the boundary of each cluster generated locally.
3) $Node_i$ joins a group G of D elements;
   // $Node_i$ joins his neighbourhood.
4) Compare cluster of $Node_i$ to other Node's clusters in the same group;
   // look for overlapping between Clusters
5) j= elect leader Node();
   // elect a node which will merge the overlapping clusters
**if** (i <> j) **then**
   Send(contour i, j);
**else**
   **if**( level > 0) **then**
      level - - ;
      Repeat 3, 4, and 5 until level=1;
   **else**
      return ($K_g$: $Node_i$'s clusters);

### B. Example of execution

We suppose that the system contains 5 Nodes (N=5), and each Node executes K-Means algorithm with different $K_i$, as it is shown in Fig 2. *Node1* executes the K-Means with *K=30*, *Node2* with *K=60*, *Node3* with *K=90*, *Node4* with *k=120*, and *Node5* with *K= 150*. Therefore each node in the system generates its local clusters. The next step consists of merging overlapping clusters within the neighbourhood. As we can see, although we started with different values of *K*, we generated only five clusters results (See Fig 2).

## V. EXPERIMENTAL RESULTS

In this section, we study the performance of $D^2CA$ Algorithm and demonstrate its effectiveness compared to BIRCH and CURE algorithms:

**BIRCH:** We used the implementation of BIRCH provided by the authors in [33]. It performs a pre-clustering and then uses a centroid-based hierarchical clustering algorithm. Note that the time and space complexity of this approach is quadratic to the number of points after pre-clustering. We set parameters to the default values suggested in [33].

**CURE:** We used the implementation of CURE provided by the authors in [34]. The algorithm uses representative points with shrinking towards the mean. As described in [34], when two clusters are merged in each step of the algorithm, representative points for the new merged cluster are selected from the ones of the two original clusters rather than all the points in the merged clusters.

**$D^2CA$**: Our algorithm is described in Section IV. The key point in our approach is to choose $K_i$ bigger than the correct number of clusters. As described at the end of Section IV, when two clusters are merged in each step of the algorithm, representative points of the new merged cluster are the union of the contours of the two original clusters rather than all points in the new cluster. This speeds up the execution time without adversely impacting on the quality of the generated clusters. In addition, our technique uses the tree topology and heap data structures. Thus, this also improves the complexity of the algorithm.

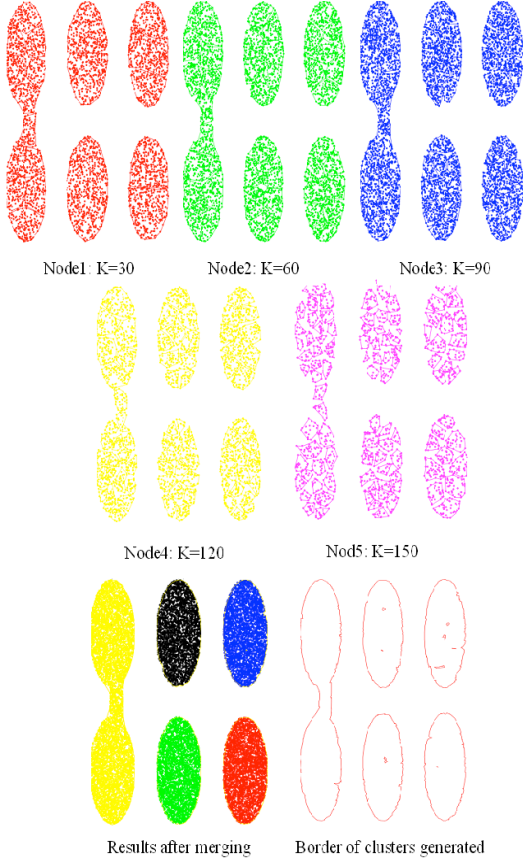

Fig 2. Distributed Dynamic Clustering Algorithm ($D^2CA$)

TABLE1
DATASETS

|  | Number of Points | Shape of Clusters | Number of Clusters |
|---|---|---|---|
| Dataset1 | 14000 | Big Oval (Egg Shape) | 5 |
| Dataset2 | 30350 | 2small Circles, 1big Circle and 2 Ovals linked | 4 |
| Dataset3 | 17080 | 4 Circles, 2Circles linked | 5 |

*A. Data sets*

We run experiments with different datasets. In this paper we use three types of datasets. These are summarised in Table 1. The number of points and clusters in each dataset is also given in Table 1. We show that $D^2CA$ algorithm not only correctly clusters the datasets, but also its execution time is much quicker than BIRCH and CURE.

*B. Quality of Clustering*

We run the three algorithms on the three datasets to compare them with respect to the quality of clusters generated and their response time. Fig 3, Fig 4 and Fig 5 show the clusters found by the three algorithms for the three datasets (dataset1, dataset2 and dataset3). We use different colours to show the clusters returned by each algorithm.

Fig 3 shows the clusters generated from the dataset1. As expected, since BIRCH uses a centroid-based hierarchical clustering algorithm for clustering the pre-clustered points, it could not find all the clusters correctly. It splits the larger cluster while merging the others. In contrast, the CURE algorithm succeeds to generate the majority of clusters but it still fails to discover all the correct clusters. Our distributed clustering algorithm successfully generates all the clusters with the default parameter settings described in section IV. As it is shown in Fig 3, after merging the local clusters, we generated five final clusters.

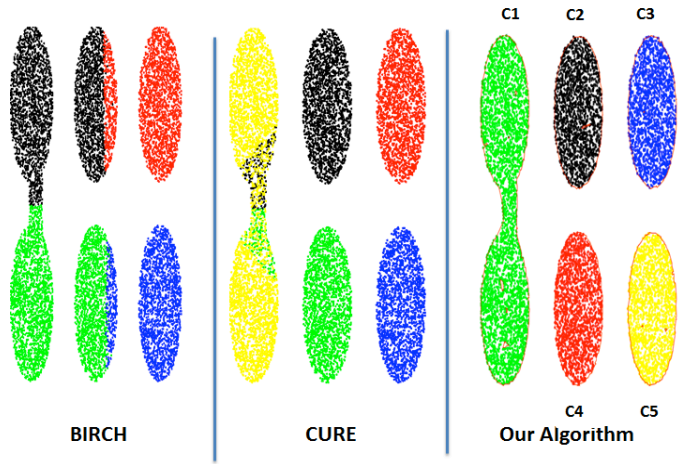

Fig 3. Clusters generated from dataset1.

Fig 4 shows the results found by the three algorithms for the dataset 2. Again, BIRCH and CURE failed to generate all the clusters, while our algorithm successfully generated the four correct clusters.

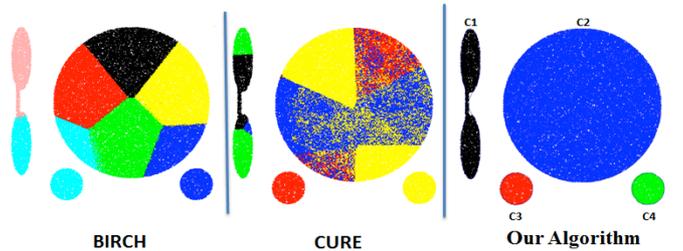

Fig 4. Clusters generated from dataset2.

Fig 5 illustrates the clustering we found from the dataset 3. As we can see BIRCH still fails to find all the clusters correctly. In contrast CURE found the 5 clusters, but not perfectly. For instance, we can see some red points in the blue cluster and some blue points in the green cluster. Our Algorithm generated the five clusters correctly and perfectly.

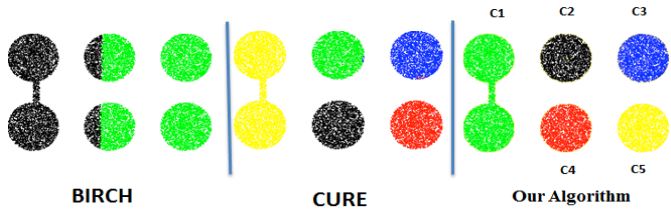

Fig 5. Clusters generated from dataset 3.

### A. Observations.

As we can see, our technique successfully generated the final clusters for the three datasets. This is due to the fact that:
- When two clusters are merged, the new cluster is represented by the union of the two contours of the two original clusters. This speeds up the execution times without impacting the quality of clusters generated
- The number of global clusters is dynamic.

### B. Comparison of $D^2CA$'s Execution Time to BIRCH and CURE

The goal here is to demonstrate the impact of using the combination of parallel and distributed architecture to deal with the limited capacity of a node in the system and tree topology to accelerate the speed of computation.

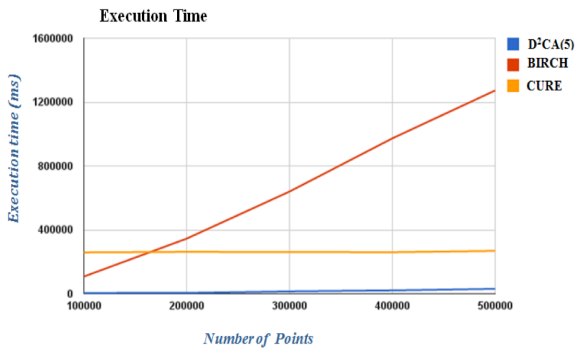

Fig 6. Comparison to BIRCH and CURE.

Fig. 6 illustrates the performance of our algorithm compared to BIRCH and CURE as the number of data points increases from 100,000 to 500,000 the number of clusters and their shapes are not altered. Thus, for our algorithm we consider the number of nodes in the system: N=5. The execution times do not include the time for displaying the clusters since these are the same for the three algorithms.

As can be seen in Fig 6, **$D^2CA$**'s execution time is much lower than CURE's and BIRCH's execution times. In addition, as the number of points increases, our execution time is nearly horizontal, while, the executions time of BIRCH increases very rapidly with the dataset size. This is because BIRCH scans the entire database and uses all the points for pre-clustering. Finally as the number of points increases the CURE's execution time is nearly horizontal, since CURE uses a sampling technique, where the size of this sample stays the same and the only additional cost incurred by CURE is the sampling procedure itself.

The above results confirm that our distributed clustering algorithm is very efficient compared to both BIRCH and CURE either in quality of the clusters generated and in the computational time.

### C. Scalability

The goal of the scalability experiments is to determine the effects of the number of nodes in the system on the execution times. The dataset contains 1000,000 points. Fig 7 shows the execution time against the number of nodes in the system. Our algorithm took only a few seconds to cluster 1000,000 points in a distributed system that contains over 100 nodes. Thus, the algorithm can comfortably handle high-dimensional data because of its low complexity.

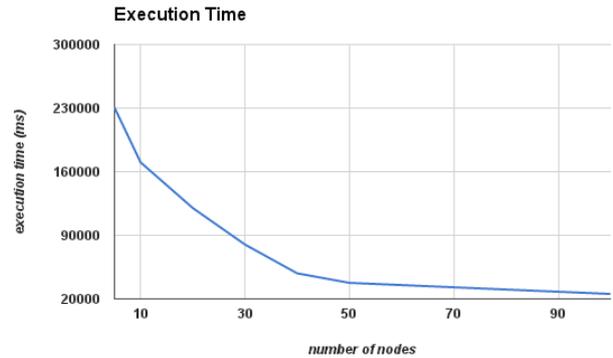

Fig 7. Scalability Experiments.

### VI. CONCLUSIONS

In this paper, we propose a new and innovative Distributed Dynamic Clustering Algorithm ($D^2CA$), to deal with spatial datasets. This approach exploits the processing power of the distributed platform by maximising the parallelism and minimising the communications and mainly the size of the data that is exchanged between the nodes in the system. Local models are generated by executing a clustering algorithm in each node, and then the local results are merged to build the global clusters. The local models are represented so that their sizes are small enough to be exchanged through the network.

Experimental results are also presented and discussed. They also show the effectiveness of $D^2CA$ either on quantity of the clusters generated or the execution time comparing to BIRCH and CURE algorithms. Furthermore, they demonstrate that the algorithm not only outperforms existing algorithms but also scales well for large databases without sacrificing the clustering quality. $D^2CA$ is different from current distributed clustering models presented in the literature, it characterises by the dynamic number of clusters generated and its efficient data reduction phase.

A more extensive evaluation is ongoing. We will intend to run experiments with various local algorithms and explore the possibilities of extending the techniques to other types of large and distributed datasets.

ACKNOWLEDGMENT

The research work is conducted in the Insight Centre for Data Analytics, which is supported by Science Foundation Ireland under Grant Number SFI/12/RC/2289.